\documentstyle[amssymb,prl,aps,epsf, multicol]{revtex}

\begin{document}

\title{Bond percolation of polymers}
\author{Manoj Gopalakrishnan$^{1,2}$, B.~Schmittmann$^{1}$ and R.~K.~P.~Zia$^{1,3}$}
\address{$^1$Department of Physics and 
Center for Stochastic Processes in Science and Engineering,\\
Virginia Tech, Blacksburg, VA 24061-0435, USA;\\
$^2$MPI f\"ur Physik komplexer Systeme, N\"othnitzer Str.~38, 01187 Dresden, Germany;\\
$^3$FB-Physik, Universit\"at Duisburg-Essen, 45117 Essen, Germany.}
\date{April 12, 2004}
\maketitle

\begin{abstract}

We study bond percolation of $N$ non-interacting Gaussian polymers of 
$\ell$ segments on a $2D$ square lattice of size $L$ with 
reflecting boundaries. Through simulations, we find the fraction of 
configurations displaying {\em no} connected cluster which span from one 
edge to the opposite edge. From this fraction, we define a critical 
segment density $\rho_{c}^L(\ell)$ and the associated critical fraction 
of occupied bonds $p_{c}^L(\ell)$, so that they can be identified as the 
percolation threshold in the $L \rightarrow \infty$ limit.
Whereas $p_{c}^L(\ell)$ is found to decrease monotonically with $\ell$ 
for a wide range of polymer lengths, $\rho_{c}^L(\ell)$ is non-monotonic. 
We give physical arguments for this intriguing behavior in terms of the 
competing effects of multiple bond occupancies and polymerization. 
\end{abstract}

\pacs{05.40.Fb, 64.60.Ak, 61.41.+e}

\vspace{-0.8cm}
\begin{multicols}{2}

Studied for many decades, the percolation problem is rich and venerable
\cite{Genref}. To this day, interesting new issues continue to surface,
for example, in the context of percolation of circles\cite{CIRCLE}, ellipses%
\cite{ELLIPSE}, random surfaces\cite{SURFACE} and polymers\cite{CORNETTE}.
Motivated by recent experiments on transport of small gas molecules across a
polymer membrane\cite{LMSZ}, we investigate a novel aspect, namely, bond
percolation with a specific correlation between the bonds. In particular, we
consider a standard model for polymers: a chain of $\ell $ segments linked
to form a random walk (RW) in space. If, instead of continuous space, we use
discrete lattices, it is natural to place the segments on the bonds. 
With certain considerations for our physical system in mind,
we study ordinary (i.e., non self-avoiding) RW's. Clearly, the locations of
the bonds occupied by one polymer are strongly correlated, except for the
$\ell =1$ case. If we place $N$ such polymers randomly on a finite
lattice and ask when the occupied bonds percolate, we face a
``correlated-bond'' percolation problem.

Before defining our problem in detail, let us devote a 
paragraph to a typical experiment. First,
a thin polymer membrane is formed by quenching a polymer melt to a
temperature far below its glass transition. To study permeation by small gas
molecules, a pressure gradient is set up across the membrane and the gas
current, in an effective steady state, is measured. The focus of the
experiments in \cite{LMSZ}, namely, how aging the membrane affects such
transport properties, serves only as a motivation for our model and
investigation here. Detailed studies of gas transport will be published
elsewhere.

For simplicity, we introduce a two-dimensional ($2D$) version of this
system. As the polymers originated from a rubbery state, we feel 
justified to neglect the complications due to self-avoidance; further, the
polymer matrix is essentially static on the time scales associated with gas
transport. Therefore, we represent each polymer as a simple RW, on a square
lattice of size $L^{2}$ (cf.~Fig.~1). 
To model a finite system with real boundaries, we
impose brick wall boundary conditions on these walks. To be specific, we
start a ``walker'' at a randomly chosen site, with each step to a 
nearest-neighbor site (a bond) modeling a segment of 
the polymer. When a walker
arrives at a boundary site, the probability for reversing direction is 1/2
(instead of 1/4 in the bulk). In this way, a bond represents a ``segment''
of the polymer of the persistence length, rather than a physical monomer. 
Note that, since the RW's are not self-avoiding, each bond may be
multiply occupied and we will use the term ``$m$-occupied'' to signify a
bond carrying $m$ segments. Meanwhile, the gas molecules are located on the
cells of our lattice and diffusion is modeled by hopping to a
nearest-neighbor cell, \emph{across} a bond. The rate for a hop across an $m$%
-occupied bond will be controlled by an activation barrier: $\exp \left(
-m\epsilon /k_{B}T\right) $, where $T$ is the temperature and $\epsilon $ is
an energy associated with jumping over a single segment.

\vspace{-0.1cm}
\begin{figure}[tb]
\epsfxsize=2.0in
\hspace{1.5cm}
\epsfbox{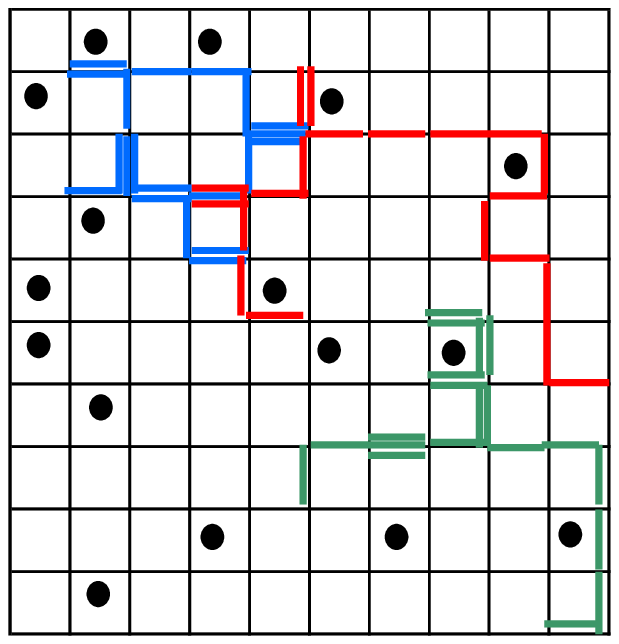}
\caption{Gas molecules diffusing from left to right 
across a $10 \times 10$ lattice, 
carrying $N=3$ polymers of length $\ell=20$.}
\end{figure}

\vspace{-0.2cm}

As the polymers are static in this model, we are dealing with driven
diffusion in a \emph{quenched} random medium. Transport through static
disordered media with random, energy barriers has been extensively studied
in the past few decades. Unlike many previous studies \cite{KIRK} of this
type, the energy barriers in our medium possess strong \emph{spatial
correlations} since the segments of a single RW are highly correlated. In
this letter, we focus on a much simpler question: In the low-$T$,
high-$\epsilon $ limit, what is the probability that there will be \emph{any}
current through the membrane? In this limit, crossing any occupied bond is
forbidden. Thus, for a $2D$ system, the quantity of interest is
precisely the probability for the occupied bonds to form a connected cluster
that spans the system in the direction \emph{transverse} to the pressure
gradient, i.e., the percolation probability in one of the directions. Of
course, the picture is more complex for a $3D$ system, on which
we will comment at the end. 

\vspace{-0.5cm}

\begin{figure}[tb]
\epsfxsize=2.2in
\epsfbox{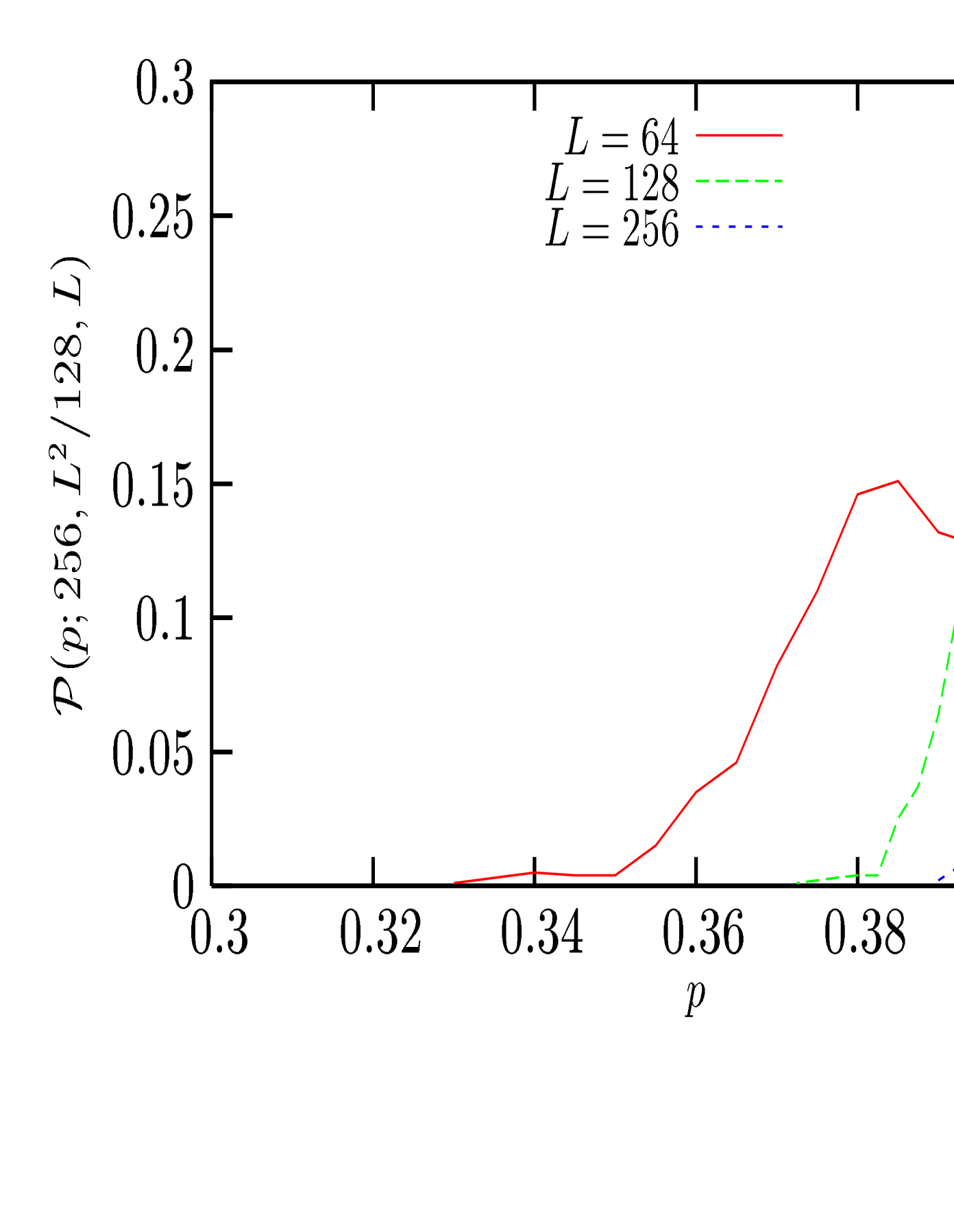}
\vspace{-1.1cm}
\caption{The probability distribution 
$\mathcal{P}$$\left( p;\ell,N,L \right)$
for the fraction of occupied bonds $p$ for $\protect\rho=1$, $\ell=256$,
and different system sizes. Inset: The mean value $\bar p$ vs.~polymer 
length $\ell $ for $\protect\rho=1/2$ and $\protect\rho=1$. 
The RMS deviation for all data points is of the order of $10^{-3}$. }
\end{figure}
\vspace{-0.3cm}

Let us briefly comment on the differences between our focus and earlier
studies which addressed this problem\cite{CORNETTE,DURR}. Previously, all
polymers were modeled as self-avoiding walks (SAW). Clearly, these pose more
challenging combinatorial questions than our RW's, so that analytic progress
is essentially impossible. In another respect, however, SAW's
are simpler: Since bonds are at most singly occupied, a given density of
polymers maps directly into a delta distribution for the probability of bond
occupation. Further, only relatively short polymers ($\ell \leq 16$ in \cite%
{CORNETTE}) were considered, focusing on singular properties near the
percolation threshold. Thus, finite size scaling techniques were invoked to
draw conclusions in the thermodynamic limit. By contrast, the samples in
permeation experiments are quite thin so
that $\ell $ and $L$ can become comparable. As a result, we will need to pay
attention to finite system sizes here. To the best of our knowledge, ours is
the first attempt to study percolation of polymers on a lattice with
multiple bond occupancies, in this particular scaling limit. 

Turning to the specifics of our problem, we place the pressure gradient in
the $x$-direction, so that the presence of a particle current translates
into the absence of a connected cluster of occupied bonds spanning the
boundaries $y=1,L$. Our question is now well posed:
Of all systems with $N$ random walks (polymers) of $\ell $ steps
(segments), placed randomly on a $L^{2}$ lattice, what is $f$, the fraction
which contains \emph{no }cluster of connected segments spanning the system
in one of the directions?

Clearly, in a gas permeation experiment, $f$ is just the fraction
of `open' configurations, allowing nonzero current. Let us emphasize that,
due to multiple occupancies, there is no simple, 1-1 relation between $p$,
the fraction of occupied bonds, and the segment density, $\rho \equiv N\ell
/(2L^{2})$ (proportional to the mass density of the sample). In
contrast, for the SAW's in \cite{CORNETTE}, $p$ is exactly $\rho $. Here, we
have a \emph{distribution} of $p$'s, a quantity to be denoted by 
$\mathcal{P}$$\left( p;\ell,N,L \right) $. 
Even in the $\ell =1$ case -- for which single
segments are randomly placed and we can easily find the probability for
a bond to be \emph{unoccupied}: $[1-1/(2L^{2})]^{N}$ -- the entire
distribution is nontrivial \cite{STIRLING}.
For arbitrary $\ell $, we are not aware of an analytic
expression for $\mathcal{P}$. However, for large $N$ and $L$, with fixed $%
\ell $, we may expect $\mathcal{P}$ to be sharply peaked, as illustrated in
Fig.~2. In all cases, we may define the average 
$\bar{p}$ $\equiv \sum p\mathcal{P}$ and use it as a control parameter 
instead of $\rho $. Of
course, their relationship is of interest. In general, we have no analytic
expression for $\bar{p}\left( \rho ;\ell ,L\right) $ though we are able to
find good estimates in some regimes \cite{THEORY}. To demonstrate this
nontrivial dependence, we show in the inset of Fig.~2 simulation results for 
$\bar{p}\left( \rho ;\ell ,L\right) $ vs $\ell $ at two different densities
and $L=128$. We find that increasing $\ell $ at fixed $\rho $
decreases the fraction of occupied bonds, since longer polymers tend to
overlap more. Note that neither the details of $\mathcal{P}$ nor its average 
$\bar{p}$ are needed for determining the main quantity of interest: $f\left(
\ell ,N,L\right)$. We raise these issues related to $p$ purely for those
familiar with the standard question of ordinary bond percolation: If a
fraction $p$ of the bonds is occupied, what is the probability for the
system to percolate?

\vspace{-0.3cm}
\begin{figure}[tb]
\epsfxsize=2.0in
\epsfbox{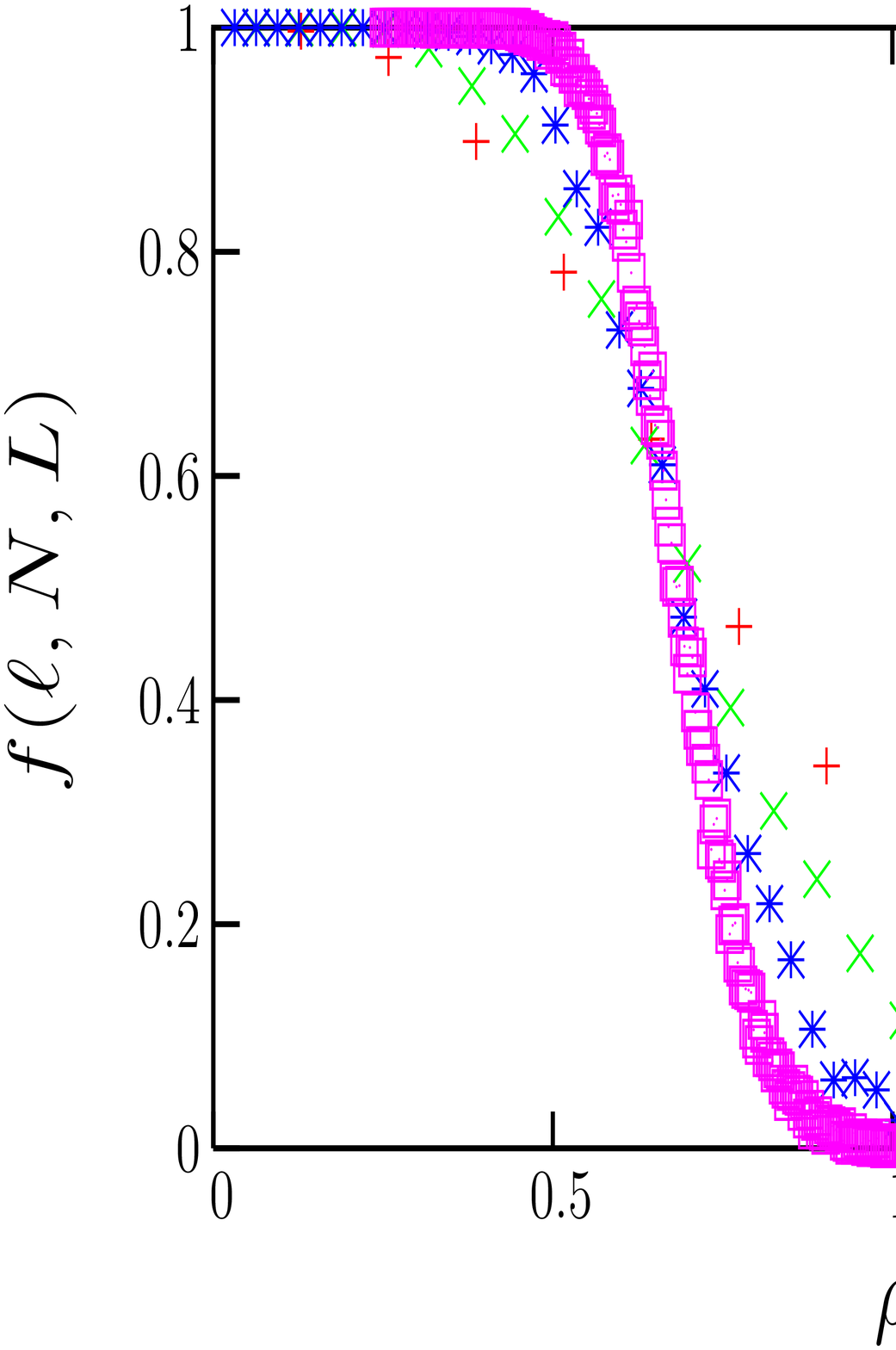}
\vspace{-0.4cm}
\caption{The percolation probability $f(\ell, N, L)$ vs.~the segment 
density $\protect\rho$ for fixed $\ell=256$ and several $L$.} 
\end{figure}

\vspace{-0.2cm}

To study $f$, we perform simulations for various $\rho \in \left[ 0.25,2.0%
\right] $ and $L\in \left[ 32,256\right] $. With a full range of $\ell $'s
and $N$'s, we find not only the typical percolation transition, but also
seemingly contradictory effects due to the correlations. Focusing on the
first aspect, we consider the data for \emph{fixed} $\ell $ but increasing $%
L $. For example, in Fig.~3, we plot $f$ against $\rho $, with $\ell =256$
and $L=32,64,128,256$. Not surprisingly, $f$ is essentially unity for small $%
\rho $ and zero for large $\rho $. In the $L\rightarrow \infty $ limit, we
expect $f$ to become a step function, vanishing for $\rho >\rho _{c}$, i.e.,
the ``percolation threshold.'' For our finite systems, we see the approach
to this behavior, namely, $f(\rho )$ falling off more steeply as $L$ 
increases. As an estimate for $\rho _{c}$, we construct the (average)
local gradients, $-\partial _{\rho }f(\rho )$, which we expect to approach 
$\delta \left( \rho -\rho _{c}\right) $ with increasing $L$. Exploiting 
$-\partial _{\rho }f$ as a ``distribution'' over 
$\rho \in \left[ 0,\infty \right] $, we compute the average 
$ -\int_{0}^{\infty }\rho \partial _{\rho }f$ and define it as the 
``critical'' $\rho _{c}^L (\ell)$ for each $L$. For the above $L$'s, 
these are respectively, 0.849, 0.754, 0.711, and 0.688. A naive
extrapolation (linear in $1/L$) leads to $\rho _{c}^{\infty }(\ell
=256)\simeq 0.67$. To express this result in the usual language of
percolation, we find $\bar{p}\left( \ell ,\rho ;L\right) $ from the
numerical data and define the ``critical'' $p$ via 
$p_{c}^L(\ell)=\bar{p}\left( \ell ,\rho =\rho _{c}^L (\ell);L\right) $.
Extrapolating to $L\rightarrow \infty $ as before yields the 
percolation threshold $p_{c}^{\infty}(\ell =256)\simeq 0\allowbreak .\,295$ 
(data for $L=32,64$ and $128$ shown in Fig.~4). 
This value is consistent with the $\ell \rightarrow \infty $
projected value for the SAW's \cite{CORNETTE}: $0.349\pm 0.001$
(based on extrapolation from $\ell \leq 16$). A more quantitative comparison
must wait for better statistics and more sophisticated analyses (e.g.,
finite size scaling) and will be reported at a later time. 

\vspace{-0.5cm}
\begin{figure}
\epsfxsize=2.0in
\epsfbox{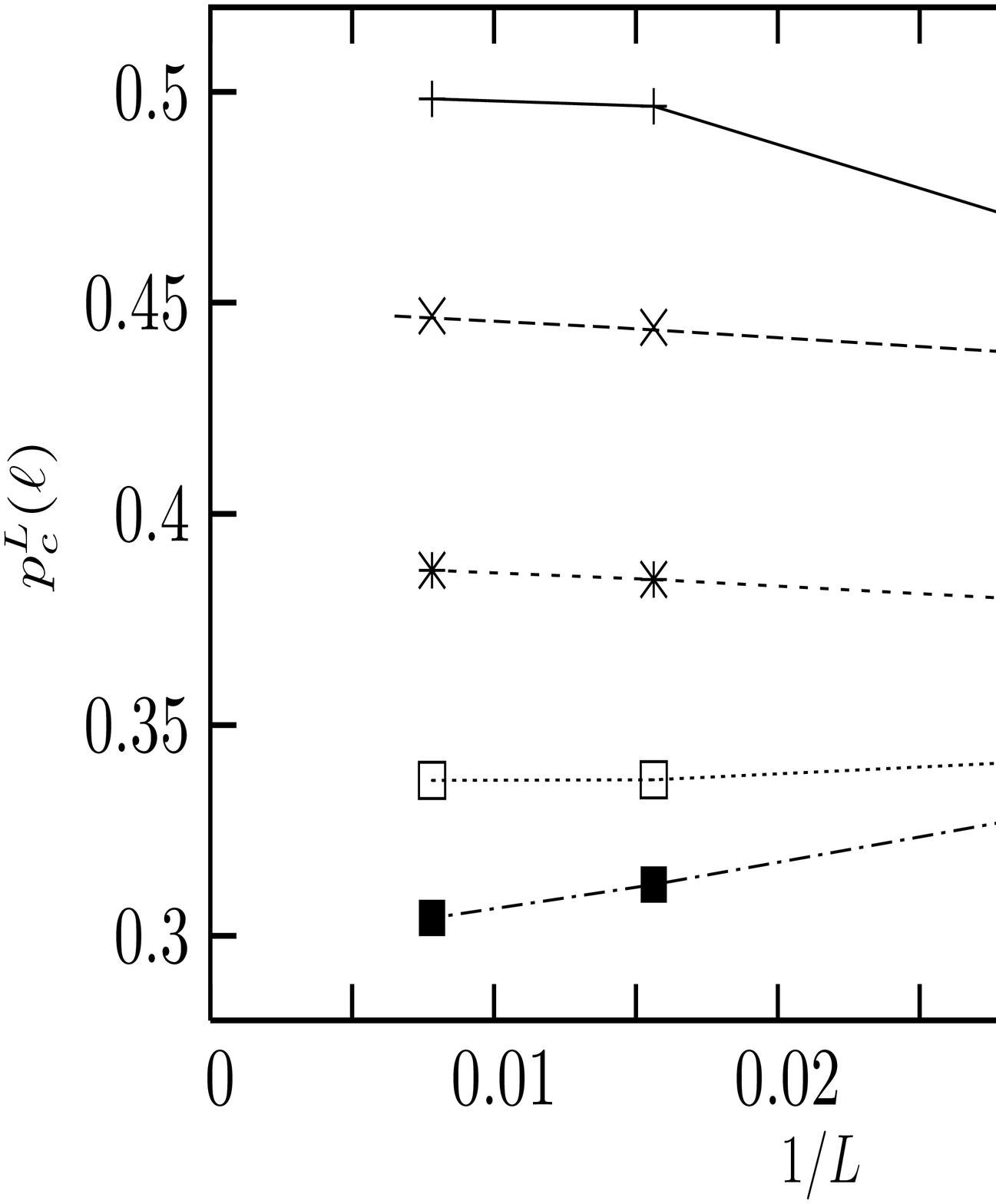}
\vspace{-0.8cm}
\caption{ The effective percolation threshold $p_{c}^L(\ell)$ vs.~$1/L$ for 
various polymer lengths $\ell$. }
\end{figure}
\vspace{-0.5cm}

Next, we turn to a second, very natural question: If we fix the \emph{density%
} of segments in a system, how will \emph{polymerizing }them, into chains of
increasing length $\ell $, affect $f(\ell ,N,L)$? 
Here, we discover a rather intriguing aspect of this
problem, illustrated in Fig.~5: For small densities, $f(\ell ,N,L)$ \emph{%
decreases }with $\ell $, but displays the opposite trend for high densities 
-- \emph{increasing} with $\ell $. For insight into these seemingly
contradictory findings, we provide some intuitive arguments. At low densities,
linking isolated segments together into longer chains clearly enhances the 
chances of spanning. As an extreme example, consider $L$ segments so that only 
one configuration can span the lattice. For $\ell =1$, at the least, all of 
them must have the same orientation and lie in the same column, so that the
probability of spanning is $o((2L)^{-L})$. Yet, if they are linked to form
a single polymer, this probability rises to $O(4^{-L})$. In contrast, at
high densities correlations have the opposite effect: if randomly
distributed segments percolate, linking them together into chains tends to
localize them and enhances overlaps. As a result, percolating paths can be
broken, and $f\left( \ell ,N,L\right) $ increases.

\vspace{-0.3cm}
\begin{figure}
\epsfxsize=1.8in
\epsfbox{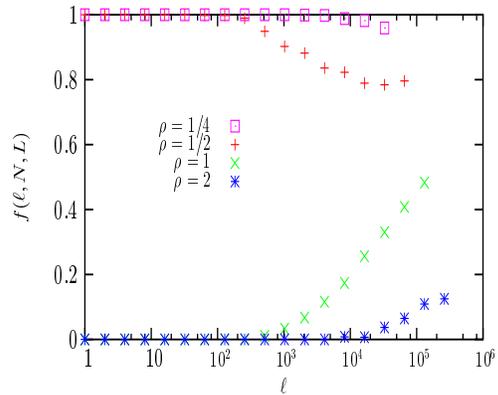}
\vspace{-0.7cm}
\caption{ The percolation probability $f(\ell, N,L)$ for fixed density 
$\protect\rho=N\ell/(2L^2)$ is plotted against $\ell$ for system size $L=128$.}
\end{figure}
\vspace{-0.3cm}

\vspace{-0.3cm}

\begin{figure}
\epsfxsize=1.8in
\epsfbox{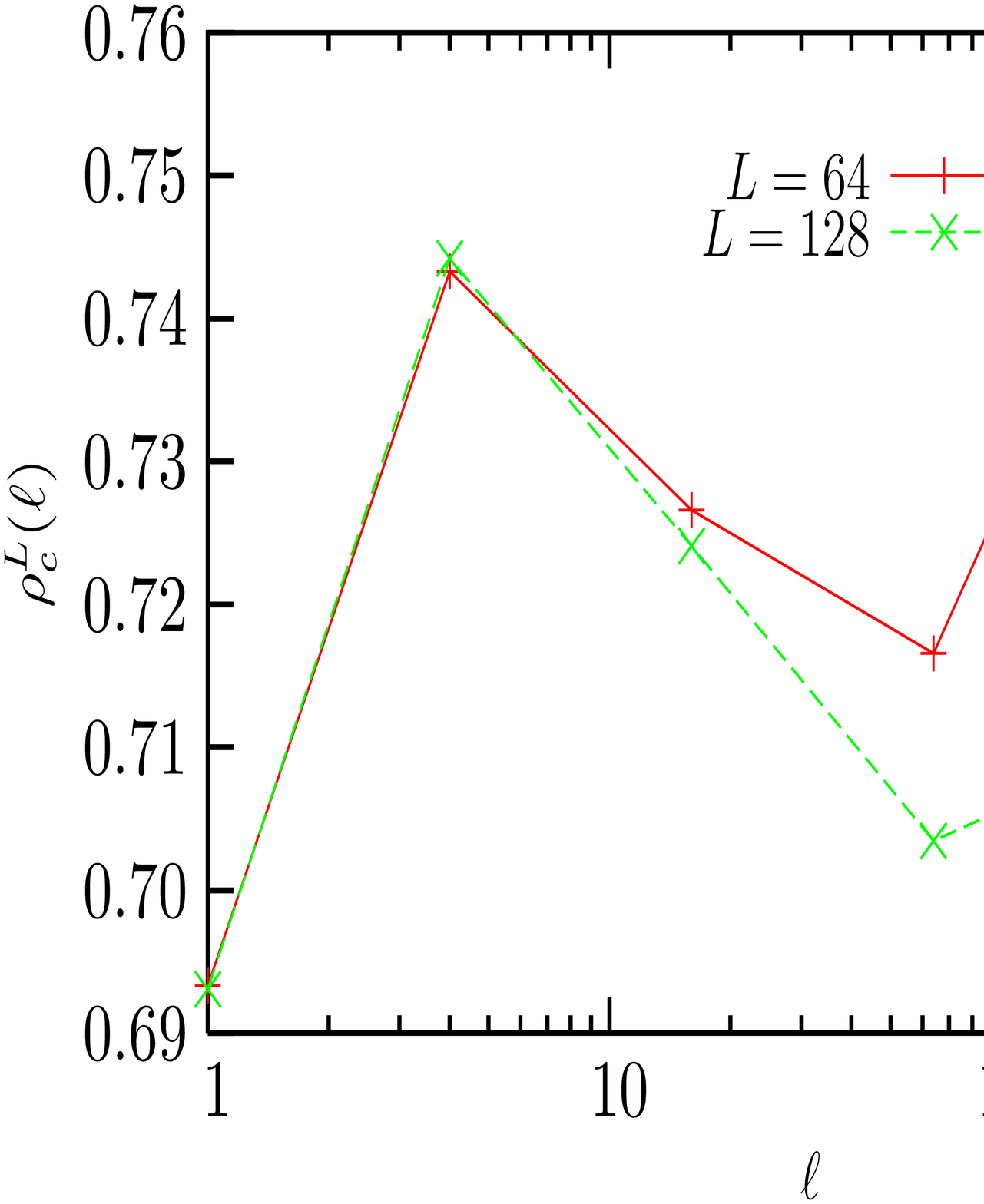}
\vspace{-0.7cm}
\caption{ The critical segment density $\protect\rho_{c}^L(\ell)$ 
vs.~$\ell$ at the percolation threshold for two system sizes. 
For $L\rightarrow \infty $, this plot may be viewed as a phase 
diagram: above the line, the polymers form a 
percolating path, while none exists below. The data points correspond to
$\ell=4^{n}$ where $n=(0,1,2,3,4)$.}
\end{figure}

\vspace{-0.3cm}

These competing effects of polymerization induce a dramatic 
$\ell$-dependence in the critical density $\rho _{c}^L$. As illustrated in Fig.~6, 
there appear to be three distinct regimes, with two extremal points for 
$\rho _{c}^L(\ell )$. Let us provide some intuitive arguments for these 
properties. 
Focusing on the first regime ($\ell \lesssim 4$), consider $\ell =1$, where 
our data suggest a critical density of $\rho_{c}^{\infty}(1)\simeq 0.69$ 
(in excellent agreement with the theoretical $\ln 2\simeq 0.693$). If we 
now generate the same segment density ($0.69$) with polymers of length $2$, 
the second segment has a probability of $1/4$ to overlap with the first one, 
so that this density will \emph{not suffice} for percolation. Thus, 
$\rho _{c}^L(2)$ must be higher than $\rho _{c}^L(1)$, in agreement with our 
observations($\rho_{c}^{L}(2)\approx$ 0.74 in simulations). 
This effect continues to dominate up to a certain value $\ell^{*}$;
from the data shown in Fig.~6, we may conclude $4\leq \ell^{*} < 16$. 
Our data indicate only a very 
weak $L$-dependence in this regime, so that we are confident that the 
$L\rightarrow \infty $ limit is essentially reached. 
Next, we consider the third regime: polymers with $\ell \gtrsim L$ 
(the last two data points in Fig.~6). Here, each polymer already occupies 
a significant fraction of the system. Thus, adding another increases $\rho$ 
noticeably, but changes bond occupations only marginally. In this sense, a 
disproportionately larger number of polymers is required to span the system, 
leading again to increasing $\rho _{c}^L$ with $\ell $. Since this regime is 
dominated by the finite size of the system, we expect it to shift to larger 
$\ell $ for bigger $L$'s.
Sandwiched between these two regimes, we observe the {\em opposite} effect,
due to polymerization: After reaching a maximum at $\ell^{*}$, the critical 
density {\em decreases}. In this regime, for reasons yet to be understood, 
the argument we presented above applies again: tieing segments together 
generates ``longer'' objects that percolate more easily. 
We conjecture that this second regime extends to infinity in the
thermodynamic limit, so that $\rho _{c}^{\infty }(\ell )$ exhibits only a 
single extremal point at $\ell^{*}$, with $4\leq \ell^{*} < 16$. 


In contrast, when expressed in terms of the bond occupancy, we observe no 
surprises: $p_{c}^L(\ell )$ is monotonically decreasing. Since we 
also expect a monotonic dependence of $p$ on $\rho $ (for fixed $\ell $), it is
all the more remarkable that the effects of correlations built into the 
segments by polymerization are so subtle to produce non-monotonic behavior
in $\rho _{c}^L(\ell )$. 

In summary, we have investigated $f(\ell ,N,L)$, the probability that 
$N$ Gaussian polymers of $\ell $ segments, randomly distributed on the
bonds of a square lattice, do not span an $L \times L$ system. Since 
the polymers are simple random walks, the bonds are frequently occupied 
by more than one segment, leading to a nontrivial correspondence between
segment density, $\rho =N\ell/(2L^2)$, and bond-occupation probability, $p$. 
For a given $\ell $, a sharp percolation threshold emerges as 
$L\rightarrow \infty $.  For finite $L$, we define ``critical'' values
$\rho _{c}^L(\ell )$ and $p_{c}^L(\ell )$. Measuring them by simulations, 
we study their behavior for various $\ell $. We find that the subtle 
interplay of polymerization and multiple occupancies leads to dramatic 
effects, especially for $\rho _{c}^L(\ell )$. Several limits, e.g., 
the extreme case of $f(\ell =2\rho L^{2},1,L)$, are analytically 
accessible and will be reported elsewhere \cite{THEORY}.

It is worthwhile to devote a few lines to comment on the experimental
implications of our work. Although our simulations were restricted to two
dimensions, we believe that our principal result, i.e., the two opposite
effects of polymerization on percolation probability (lowering $p_{c}$ on
one hand, while making more free bonds available via multiple bond
occupancies on the other) should be present in higher dimensions also.
Percolation-related transitions in permeation properties have been observed
experimentally, eg. in \cite{PERC_EXP} where a jump in the gas current is
seen when the volume fraction of the flexible component of the polymer is
increased beyond a threshold value. It would be interesting if the
non-monotonic variation of the critical density with polymer length could be
tested experimentally.

This work was supported in part by NSF grant DMR-0088451 and the Alexander 
von Humboldt Foundation. We acknowledge H.~Hilhorst, R.~Zallen, E.~Marand, 
O.~Stenull, L.~Sch\"{a}fer and R.~Kree for fruitful discussions. We thank 
H.W.~Diehl for his hospitality at the University 
of Duisburg-Essen, where some of this research was carried out.

\end{multicols}
\end{document}